\documentclass[twocolumn,prb,showpacs,preprintnumbers,amssymb]{revtex4}

\usepackage{graphicx}
\usepackage{dcolumn}
\usepackage{bm}
\usepackage{color}
\usepackage{multirow}

\begin{document}

\title{Superconducting properties of K$_{1-x}$Na$_x$Fe$_2$As$_2$ under pressure}

\author{V.\ Grinenko} \email{v.grinenko@ifw-dresden.de}
\affiliation{Leibniz-Institute for Solid State and Materials Research, IFW-Dresden, D-01171 Dresden, Germany}
\author{W. Schottenhamel}\affiliation{Leibniz-Institute for Solid State and Materials Research, IFW-Dresden, D-01171 Dresden, Germany}
\author{A.U.B.\ Wolter}\affiliation{Leibniz-Institute for Solid State and Materials Research, IFW-Dresden, D-01171 Dresden, Germany}
\author{D.V.\ Efremov}
\affiliation{Leibniz-Institute for Solid State and Materials Research, IFW-Dresden, D-01171 Dresden, Germany}
\author{S.-L.\ Drechsler} 
\affiliation{Leibniz-Institute for Solid State and Materials Research, IFW-Dresden, D-01171 Dresden, Germany}
\author{S.\ Aswartham\footnote{Dept of Physics and Astronomy, University of Kentucky, Lexington, USA}}
\affiliation{Leibniz-Institute for Solid State and Materials Research, IFW-Dresden, D-01171 Dresden, Germany}
\author{M.\ Roslova}
\affiliation{Leibniz-Institute for Solid State and Materials Research, IFW-Dresden, D-01171 Dresden, Germany}
\affiliation{Lomonosov Moscow State University, GSP-1, Leninskie Gory, Moscow, 119991, Russian Federation}
\author{I.\ V.\ Morozov}
\affiliation{Leibniz-Institute for Solid State and Materials Research, IFW-Dresden, D-01171 Dresden, Germany}
\affiliation{Lomonosov Moscow State University, GSP-1, Leninskie Gory, Moscow, 119991, Russian Federation}
\author{M. Kumar}
\affiliation{Leibniz-Institute for Solid State and Materials Research, IFW-Dresden, D-01171 Dresden, Germany}
\author{S.\ Wurmehl}\affiliation{Leibniz-Institute for Solid State and Materials Research, IFW-Dresden, D-01171 Dresden, Germany}
\affiliation{
}
\author{B.\ Holzapfel}
\affiliation{Leibniz-Institute for Solid State and Materials Research, IFW-Dresden, D-01171 Dresden, Germany}
\affiliation{Karlsruhe Institute of Technology (KIT),
Hermann-von-Helmholtz-Platz 1, 76344 Eggenstein-Leopoldshafen, Germany}
\author{B.\ B\"uchner}\affiliation{Leibniz-Institute for Solid State and Materials Research, IFW-Dresden, D-01171 Dresden, Germany}
\affiliation{Institut f\"ur Festk\"orperphysik, TU Dresden,
D-01062 Dresden, Germany}
\author{E. Ahrens}
\affiliation{Institut f\"ur Anorganische Chemie, TU Dresden, D-01062 Dresden, Germany}
\author{S.\ I.\ Troyanov}
\affiliation{Lomonosov Moscow State University, GSP-1, Leninskie Gory, Moscow, 119991, Russian Federation}
\author{S. K\"ohler}
\affiliation{Goethe University, Max-von-Laue-Strasse 1, 60438, Frankfurt am Main}
\author{ E. Gati}
\affiliation{Goethe University, Max-von-Laue-Strasse 1, 60438, Frankfurt am Main} 
\author{S. Kn\"oner}
\affiliation{Goethe University, Max-von-Laue-Strasse 1, 60438, Frankfurt am Main}
\author{N. H. Hoang} 
\affiliation{Goethe University, Max-von-Laue-Strasse 1, 60438, Frankfurt am Main}
\author{M. Lang} 
\affiliation{Goethe University, Max-von-Laue-Strasse 1, 60438, Frankfurt am Main}
\author{F. Ricci} \email{fabio.ricci@aquila.infn.it}
\affiliation{CNR-SPIN and Dipartimento di Scienze Fisiche e Chimiche, Universit\`a degli Studi dell'Aquila, Via Vetoio 10, I-67100 L'Aquila, Italy}
\author{G. Profeta}
\affiliation{CNR-SPIN and Dipartimento di Scienze Fisiche e Chimiche, Universit\`a degli Studi dell'Aquila, Via Vetoio 10, I-67100 L'Aquila, Italy}

\date{\today}

\begin{abstract}
The effect of hydrostatic pressure 
and partial Na substitution on the normal-state properties and the superconducting transition temperature
 ($T_c$) of K$_{1-x}$Na$_x$Fe$_2$As$_2$ single crystals were investigated. 
It was found that a partial Na substitution 
leads to a deviation from the standard  $T^2$ Fermi-liquid behavior in the temperature dependence of the normal-state resistivity. It was demonstrated that non-Fermi liquid like behavior of the resistivity for 
K$_{1-x}$Na$_{x}$Fe$_2$As$_2$ and some KFe$_2$As$_2$ samples can be explained by disorder effect in the multiband system with rather different  quasiparticle effective masses. Concerning the superconducting state our data support the presence of 
 a shallow minimum around 2 GPa in the pressure dependence of $T_c$ for stoichiometric KFe$_2$As$_2$. 
The analysis of $T_c$ in the K$_{1-x}$Na$_{x}$Fe$_2$As$_2$ at pressures below 1.5 GPa showed, 
that the reduction of $T_c$ with Na substitution follows the 
Abrikosov-Gor'kov law with the 
critical temperature $T_{c0}$ of the clean system 
(without pair-breaking) which 
linearly depends on the pressure.
Our observations, also, suggest that $T_c$ of K$_{1-x}$Na$_x$Fe$_2$As$_2$ is nearly independent of the lattice compression produced by the Na substitution.
Further, we theoretically analyzed the behavior of the band structure under pressure within the generalized gradient approximation 
(GGA). A qualitative agreement between the calculated and the recently in de Haas-van Alphen 
experiments \cite{Terashima2014} measured pressure dependencies of the Fermi-surface cross-sections has been found. 
These calculations, also, indicate that the observed minimum around 2~GPa in the pressure dependence of $T_c$ may occur 
without a change of the pairing symmetry.
\end{abstract}


\maketitle

\section{Introduction}
Recently, KFe$_2$As$_2$ (K-122) has attracted special attention due to several unusual physical properties such as 
a very large quasiparticle effective mass \cite{Terashima2014, Terashima2013, Hardy2013, Fukazawa2011,Grinenko2014} and superconductivity (SC) with nodes in the superconducting gap \cite{Hardy2013, Fukazawa2011,Grinenko2014,Abdel-Hafiez2013,Reid2012,Okazaki2012}. 
Such a large mass enhancement is usually ascribed to 
the presence of 
pronounced correlation effects.\cite{{Kotliar2011}} However, in the present case the nature and 
the strength of these correlation effects as well as the symmetry of the 
superconducting order parameter are still under debate. Some experiments 
were interpreted in favor of $s_{\pm}$ -wave SC  
with accidental nodes \cite{Okazaki2012,Hardy2013a,Kittaka2014,Watanabe2013} 
whereas 
other favor $d$- wave SC \cite{Reid2012,Grinenko2014,Abdel-Hafiez2013,Hashimoto2010,Kim2014}. 
Based on their theoretical calculations the authors of Ref.\ \onlinecite{Ikeda2010} argue that the electronic correlations 
are enhanced in the strongly hole-overdoped K-122 as
compared to lower hole doping levels in other 122 systems due to
the changes in the Fermi-surface topology. This might 
affect the low-energy spin excitation spectra and cause 
an increasing density of states (DOS) at the Fermi energy $E_F$. 
However, this calculation 
underestimates the correlation effect found experimentally. 
Recently, it was proposed that the unexpected 
large effective mass enhancement of the order of 9 to 10 could be related to
the proximity of an orbital-selective Mott transition 
for K-122 \cite{Hardy2013}.
The relation between the enhanced correlation effects and the
SC remains still unclear. Therefore, 
tuning of the physical properties of K-122 by external or chemical pressure 
and disorder can provide more insight 
into the pairing symmetry and the present correlation effects. 

Recently, several studies
of the physical properties of K-122 \cite{Budko2012, Budko2013,Burger2013,Tafti2013,Tafti2014,Taufour2014,Terashima2014} 
and its sister compounds Cs-122 \cite{Sasmal2008, Wang2014, Tafti2014}, and 
Rb-122 \cite{Bukowski2010, Shermadini2010, Shermadini2012, Zhang2014} under external pressure have been reported. 
F.F.\ Tafti {\it et al.} observed that the 
superconducting critical temperature $T_c$ of K-122 at first monotonically 
decreases with increasing hydrostatic pressure, then reaches  
a minimum at $P\sim$ 1.7 GPa (we adopt from Ref. \onlinecite{Tafti2013} referring $P=1.7$ GPa as critical pressure), and finally
shows a reversal behavior \cite{Tafti2013,Tafti2014}. 
The value of the critical pressure is nearly independent 
of the amount of the disorder in K-122 but it is reduced in the unsubstituted 
clean Cs-122 \cite{Tafti2014}. However, V.\ Taufour {\it et al.}, showed 
that the pressure dependence of $T_c$ for K-122 
is sensitive to the pressure medium \cite{Taufour2014}, and that for the best 
pressure (hydrostatic) conditions the reversal behavior of $T_c$ is replaced by 
a broad minimum followed by a nearly pressure independent $T_c$ value above 
2.5~GPa. From the almost pressure independent behavior of the Hall coefficient 
extrapolated to $T$ = 0 as well as from the residual resistivity, 
it was concluded that 
the topology of the Fermi surface does not change across the critical
pressure.\cite{Tafti2013, Tafti_p} This point of view 
was, also, supported by recent de Haas-van Alphen (dHvA) 
measurements under pressure, where at least two small but 
accessible Fermi surface 
sheets (FSS), namely the smallest cylinder $\alpha$ around the $\Gamma$-point 
and the propeller blade-like FSS $\epsilon$ around the X point of the Brillouin 
zone (BZ), do not show a Lifshitz transition up to 2.47~GPa.\cite{Terashima2014} 
The evolution of the Fermi surface is accompanied by a reduction of the effective quasi-particle 
masses without any special feature at the critical pressure. 
The reduction of the effective masses and
the decrease of $T_c$ up to the pressure about 2~GPa can be qualitatively explained by a reduction of coupling constants.  
In order to explain the $T_c$ minimum it was 
supposed that at the critical pressure a change of the superconducting
gap symmetry from $d$- wave to $s_{\pm}$- wave 
\cite{Tafti2013, Tafti2014} or from nodal $s_{\pm}$ -wave to a nodeless 
one \cite{ Terashima2014} should occur. The fast increase of the  
inelastic scattering rate at the critical pressure obtained from 
the transport measurements supports 
such a hypothesis.\cite{Tafti2014} Alternatively, V.\ Taufour {\it et al.} proposed that a 
change of symmetry under pressure may not occur at all.\cite{Taufour2014} 
These authors empirically supposed that the critical 
pressure could be a consequence of anisotropic pressure derivatives, 
since the pressure derivatives along 
the $a$ axis and the $c$ axis are
negative and positive, respectively. In this case, the critical pressure 
corresponds to the critical pressure in the $ab$ plane
above which the slope of $T_c(P)$ 
depends on the value of 
$\partial T_c/\partial P_c >0$, only, whereas 
$\partial T_c/\partial P_{ab} \equiv 0$ for $P\geq 2$~GPa. These authors
have also shown quantitatively that the behavior of the upper 
critical fields in crossing the critical pressure 
can be accounted for by 
$k_z$ modulations of the superconducting gap, only. 

The amount and type of disorder provide additional possibilities
to tune the superconducting properties. Recently, the effect of Na substitution
on the superconducting and the normal-state properties of 
K-122 was investigated.\cite{Abdel-Hafiez2013, Grinenko2014} It was shown that the temperature 
dependence of the specific heat of K$_{1-x}$Na$_x$Fe$_2$As$_2$ single crystals 
can be explained by multiband $d$-wave SC. The disorder induced by 
the Na substitution suppresses the small superconducting gaps and leads to
gapless SC with a large residual DOS. 
The specific heat jump scales approximately with a power-law, $\Delta C_{\rm el} \propto T_c^{\beta}$, 
with $\beta \approx 2$ determined by the impurity scattering rate, in 
contrast to most iron-pnictide superconductors described by
the remarkable Bud'ko-Ni-Canfield (BNC) scaling $\Delta C_{\rm el} \propto T^3$.
\cite{Budko2013,Budko2009} However, the Na substitution might
also produce chemical pressure which could additionally affect $T_c$. 

In this work, we show that a partial Na substitution produces 
non-hydrostatic chemical pressure with 
a larger compression along the $c$ axis.
Thereby the netto (total)
effect of the chemical pressure on $T_c$ is strongly reduced due to 
anisotropic pressure derivatives with different signs along the 
$c$ and $a$($b$) axis. Therefore, the pressure dependence of 
$T_{c0}$ (without pair-breaking) and the 
corresponding coupling constant are nearly independent of
the Na substitution. 
Furthermore, we present a theoretical analysis of the band structure of KFe$_2$As$_2$ under 
pressure within the first-principles density functional theory in the generalized gradient approximation (GGA). 
The calculations indicate that the minimum in 
the pressure dependence of $T_c$ can be qualitatively explained by a
relatively small non-monotonic variation of the density of states (DOS) 
at the Fermi level. In general, our results suggests that 
the superconductivity in K-122 is driven by a single leading band.\\

\section{Experimental}
K$_{1-x}$Na$_{x}$Fe$_{2}$As$_{2}$ single crystals 
with a typical mass of about 1-2~mg and several mm in-plane 
dimensions were grown by the self-flux method.
The compositions and the phase purity of the investigated
samples were determined by an EDX analysis in a scanning
electron microscope, and by x-ray analysis; for details 
see Refs. \onlinecite{Abdel-Hafiez2013, Grinenko2014}. 
The two stoichiometric KFe$_2$As$_2$ samples were 
prepared using different fluxes: S1 from a FeAs flux and S2 from 
a KAs one, for details see Refs.\ \cite{Abdel-Hafiez2012, Grinenko2013}.   
Additionally, we performed single crystal diffraction 
measurements at room temperature and at $T$ = 100 K. 
The crystallographic parameters were extracted from single crystal diffraction data.
The room temperature single crystal X-ray diffraction data were collected on a Bruker Kappa APEX-II CCD diffractometer using Mo-K$\alpha$ radiation. 
$\phi$- and $\omega$-scans were recorded with an increment of 0.3,  integration
 and corrections for oblique incidence and polarization  
 were performed within SAINT \cite{SAINT+}. A multiscan absorption correction 
 was  applied using SADABS \cite{Sheldrick2018}. Structure solutions and 
 refinements were  done with the JANA2006 package \cite{Petricek2006}. 
 Single crystal X-ray diffraction analysis at $T$ = 100 K was carried out with 
 an IPDS (STOE) diffractometer using graphite monochromated Mo-K$\alpha$ 
 radiation. Numerical absorption corrections were applied for both data 
 sets with transmission coefficients of $T_{min}/T_{max}$ = 0.0077/0.3052 
 and 0.0495/0.06333, respectively. The structures were solved and anisotropically 
 refined with the SHELX program package.\cite{SHELXL97} The relative Na/K contents were included 
 into the refinements as an independent parameter (x). 

The temperature dependence of the electrical resistivity was measured by
the standard four-contact method in the Quantum Design Physical Property 
Measurement System (PPMS). The temperature dependent, zero-field-cooled (ZFC) 
magnetization under pressure was 
measured in a Quantum Design MPMS-SQUID magnetometer down 
to 1.8~K in a magnetic field of 10 Oe. The
samples position was 
not perfectly fixed during the magnetization measurements under pressure. 
Therefore, in some cases samples could be tilted with 
respect to the applied field direction after the application of the pressure. 
This affects the 
amplitude of the superconducting diamagnetic signal (due to a variation 
in the demagnetization factor) as can be seen in 
Fig.\ \ref{mag_all}. However, the defined $T_c$ values are not affected 
considerably, since the measurements were performed at low applied magnetic 
field as compared to the critical fields of the investigated samples.

For the DC magnetization measurement under pressure,
we used a commercial pressure cell easyLab Mcell 10 tailored for a MPMS of Quantum Design for the pressure measurements up to 1~GPa and a homemade pressure cell (HPC) for the pressure measurements up to 
3~GPa. The HPC has been constructed for a commercial MPMS. 
Its cell design is mostly related to a diamond anvil cell (DAC), in which two opposing cone-shaped ceramic anvils compress a gasket that serves as the sample
chamber.\cite{Alireza2009,Tateiwa2012,Tateiwa2013} 
A supplied uniaxial pressure will be transformed into uniform hydrostatic pressure by using Daphne7373 oil 
as transmitting medium. The cell can be equipped with different types of sample chambers, for which the sample space varied between either 0.7 mm in diameter and 0.4 mm in depth, 
or on the other hand 0.3 mm in diameter and 0.2 mm in depth. Using these 
different chambers a maximum hydrostatic pressure 
up to 3~GPa and 6~GPa can be generated, respectively. Each mechanical part 
of the pressure cell, except for the anvils, is made from a very low-magnetic CuBe alloy, which allows
measurements at a weak background signal of the cell itself. \cite{fn_pressure} Additionally, the magnetization data have been corrected for the contribution of the background signal by subtracting the extrapolated (below $T_c$) background signal obtained at $T > T_c$. The pressure was determined at low temperatures using superconducting Sn in 
case of the Quantum Design pressure cell and Pb in 
case of the homemade cell as a low-temperature pressure gauge. 

For most of the measurements the pressure dependence 
of $T_c$ was linear up to 1.5 GPa (see Fig. \ref{Tc_P}). However, using the easyLab Mcell 10 
pressure cell in some cases we observed deviations from a 
linear $T_c(P)$ dependence at pressures below 0.1~GPa. 
To verify the origin of this behavior at low pressures, 
we performed additional measurements in a He-gas CuBe pressure cell up to the maximum pressure of 0.5~GPa. In the experiment a CuBe cell connected to a He-gas compressor was used for finite $P$ measurements. \cite{Institute} An In sample was used for an in-situ determination of the pressure. \cite{Jennings1958} 
The use of 4-He as a pressure-transmitting medium ensures truly hydrostatic pressure conditions at low temperatures. \cite{Gati2012,Taufour2014} The susceptibility data have been corrected for the contribution of the sample holder, including the pressure cell, which was determined independently. In the case of the experiments with He as a pressure medium the non-linear behavior in the $T_c(P)$ dependence at low pressures was not observed. Therefore, we may conclude that the observed deviation from the linear $T_c(P)$ dependence is related to deviation from the hydrostatic conditions. In this work, we consider the data with a linear $T_c(P)$ dependence at low pressure, only, assuming that in this case the measurements where performed under hydrostatic pressure conditions.\\

\section{Results}
\subsection{Resistivity  and Magnetization}

\begin{figure}
\includegraphics[width=\columnwidth,clip]{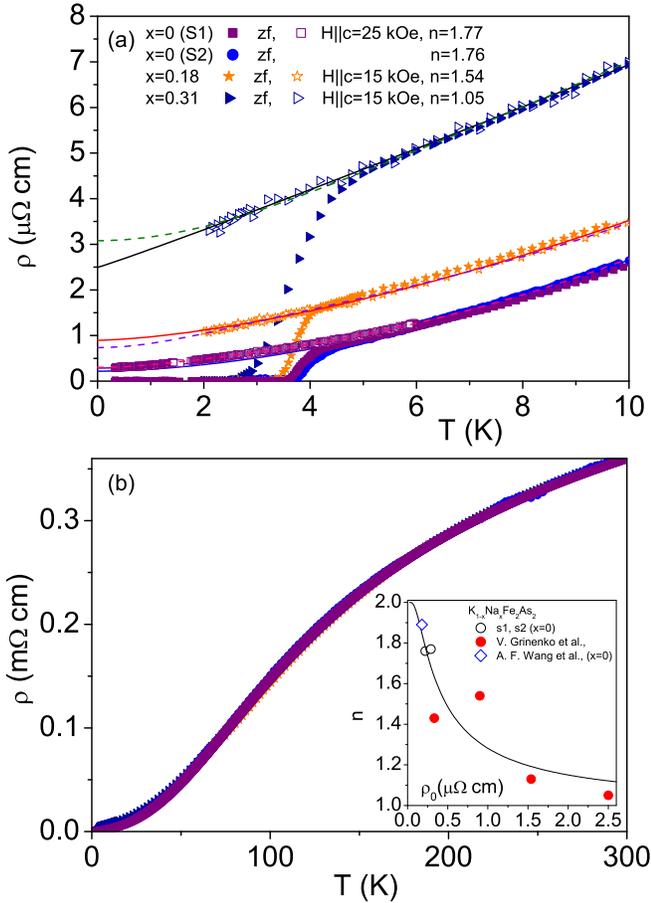}
\caption{(color online) The $T$ dependence of the resistivity of K$_{1-x}$Na$_x$Fe$_{2}$As$_{2}$ at zero pressure. 
The data 
for $x$ = 0.18 and  
$x$ = 0.31 are taken from the supplementary information in Ref.\ \onlinecite{Grinenko2014}.
 (a) The temperature range below 10 K, symbols - experimental data, solid lines - single band fit Eq.\ref{Eq1}, 
 dashed lines - multiband fit Eq.\ \ref{Eq2}. (b) The temperature range up to 300 K.
  Inset: the dependence of the exponent $n$ vs.\ the residual resistivity $\rho_{0}$, see Eq.\ (1). 
 The data for K$_{1-x}$Na$_x$Fe$_{2}$As$_{2}$ crystals ($x\neq 0$) are taken from the supplementary information in Ref.\  \onlinecite{Grinenko2014}, 
 for $x$ = 0 (open rhombus) from Ref.\ \onlinecite{Wang2014}. The solid line is a guide-to-the-eye.}
\label{res}
\end{figure}

\begin{figure*}
\includegraphics[width=16cm]{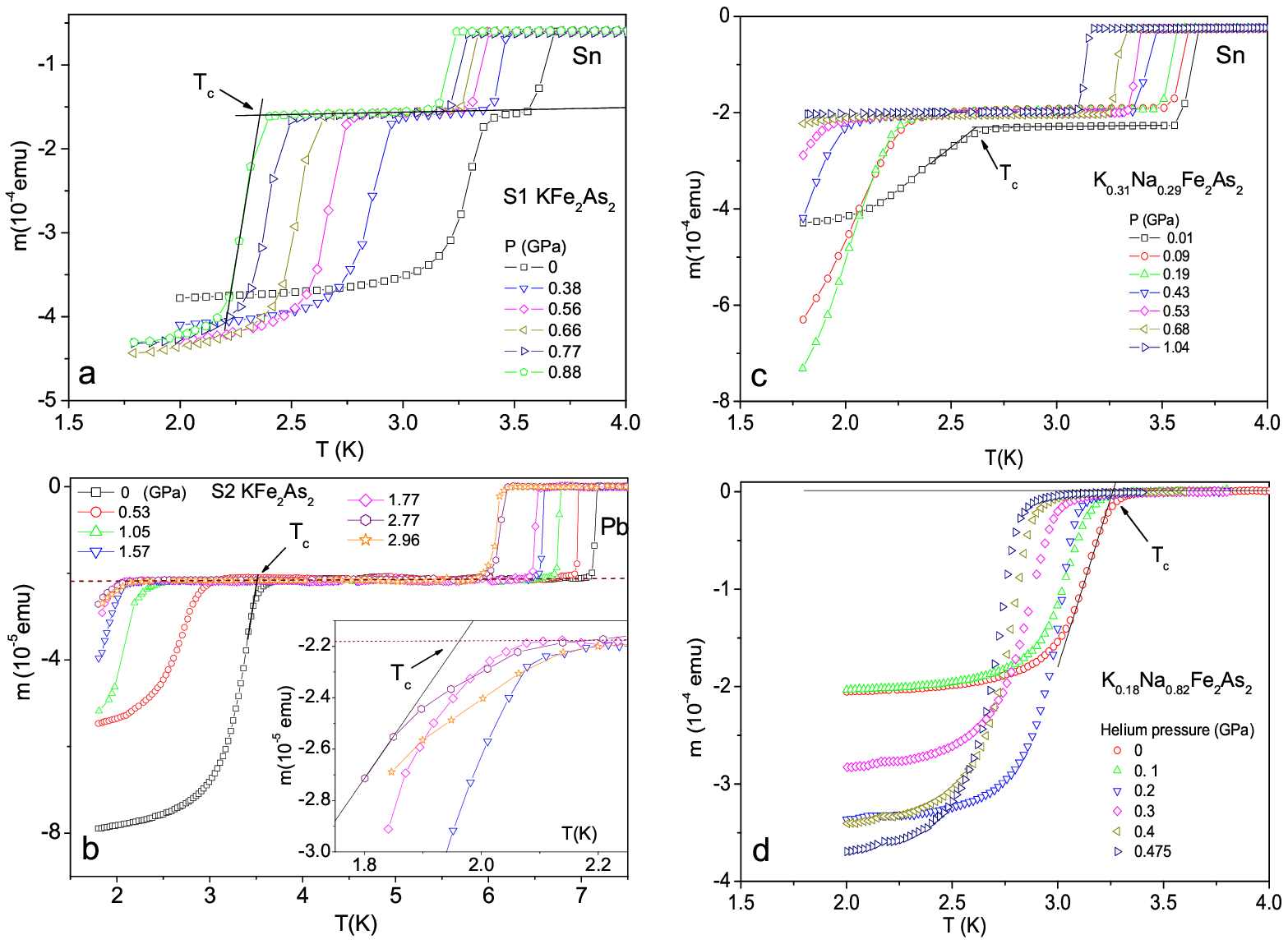}
\caption{(color online) The $T$ dependence of the magnetization of K$_{1-x}$Na$_x$Fe$_{2}$As$_{2}$ measured in 10 Oe at 
different applied pressures. The second superconducting transition at higher temperature (Figs. a, b, and c) corresponds to a low-temperature pressure gauge such as Sn or Pb. a) $x$ = 0 (S1 sample) measured in a 
Quantum Design pressure cell. b) $x$ = 0 (S2 sample) measured in a home-made pressure cell. 
Inset: zoom to the data at high pressure. c) Sample with $x = 0.31$ measured in a Quantum Design pressure cell, d) Sample with $x = 0.18$ measured in a 
He gas pressure cell. The data in Fig. b,d were corrected for the background signal.}
\label{mag_all}
\end{figure*}

Fig.\,\ref{res} shows the $T$ dependence of the electrical resistivity 
$\rho(T)$ of K$_{1-x}$Na$_x$Fe$_{2}$As$_{2}$ single crystals. To remove the uncertainty in the absolute value of $\rho(T)$ between our transport data and the published one in literature associated with geometric factors, 
we used a normalization procedure as proposed in Ref. \onlinecite{Reid2012}. In this way, the resistivity data $\rho(T)$ of each sample were multiplied by a factor $\langle\rho(300$K$)\rangle/\rho(300$K$)$, where  $\rho(300$K$)$ is the measured resistivity of the sample at $T$ = 300 K and  $\langle\rho(300$K$)\rangle$ = 0.36 m$\Omega$cm is our average resistivity value over all of our samples
at $T$ = 300 K. Below we show only normalized  $\rho(T)$ data. 
To fit the data below 15~K, we used the standard single-band expression 
\begin{equation}
\rho(T)=\rho_0+AT^n.
\label{Eq1}
\end{equation}
The obtained residual resistivity $\rho_0$ monotonically increases with 
the Na doping level as shown in Fig.\,\ref{res}a. In turn, 
$T_c$ decreases with the increasing $\rho_0$ 
in accord with the presence of the disorder induced pair-breaking.\cite{Grinenko2014} 
We have found that the exponent $n$ is 
always below 2: $n \approx 1.8$ for our stoichiometric 
K-122 samples and it decreases to $n \approx 1$ for 
the single crystals with a high Na substitution 
level. In the case of single-band materials such a 
deviation from a $T^2$ dependence would signal a non-Fermi liquid (nFL) behavior. For example, 
the deviation from the FL behavior in the strongly disordered K-122 has 
been interpreted as 
evidence for quantum criticality \cite{Dong2010}. However, in a multiband system, which K-122 is, 
the deviation from the $T^2$ behavior can be observed 
even in the case when all $i$ bands behave like $\rho(T)_i=\rho_{0i}+A_iT^2$. 
For the resulting resistivity of the multiband system we have: 
\begin{equation}
\rho(T)=[\sum_i(1/\rho_i(T))]^{-1}.
\label{Eq2}
\end{equation}
Strictly speaking, in the case of K-122 a four (five)  
band-model should be considered, if one neglects (takes into account) the  
small 3-dimensional band near the Z point of 
the BZ (see Fig.\,\ref{FS_exp}) with 8(10) independent parameters. To demonstrate 
that the data can be fitted by Eq.\,\ref{Eq2} 
we considered for the sake of simplicity an effective two-band model as shown in Fig.\,\ref{res}a.
(The details of the fitting parameters are given in Table \ref{tab:2}.)
As can be seen, a two-band analysis 
results in quite different $A_i$ values. The value of the coefficient $A_1$ is close to the 
values reported for K-122 compounds obtained using the single-band Eq.\  
\ref{Eq1}. \cite{Taufour2014,Hashimoto2010}. However, the value of the coefficient $A_2$ is about one 
order of magnitude larger than that of $A_1$. This difference can be understood 
if we consider that  $A\propto n/N_0(0)^2\langle \upsilon_{ab}^2 \rangle \omega^{*2}$, 
where $n$ is the density of the carriers, $N_0(0)$ is the bare density 
of states (DOS) at the Fermi energy, $\upsilon_{ab}$ is the bare Fermi 
velocity in the $ab$ plane, and $\omega^{*}$ is the frequency which 
determines the magnitude of the frequency dependent term in the imaginary 
part of the self-energy and contains all of the many-body effects. 
\cite{Jacko2009} Then, the ratio $A_2/A_1 \sim 10$ can be explained by 
the ratio of $\omega_1^{*2}/\omega_2^{*2}$, only, by taking into 
account rather different band dependent quasiparticle effective masses 
for K-122 \cite{Terashima2013, Hardy2013}. Therefore, the obtained 
parameters (see Table \ref{tab:2}) have reasonable values for the 
K$_{1-x}$Na$_x$Fe$_{2}$As$_{2}$ system 
under consideration which suggests that the nominally nFL behavior of the 
total $\rho(T)$ can be explained by a multiband effect with all bands in a FL state. 

 \begin{center}
\begin{table}
		\vspace{0.5cm}
		\begin{center}
		\scalebox{1}{
		\begin{tabular}{|c|c|c|c|c|}
		\hline
		 x & $\rho_{01}$($\mu\Omega$cm) & $\rho_{02}$($\mu\Omega$cm)  & $A_1$($\mu\Omega$cm/K$^2$)& $A_2$($\mu\Omega$cm/K$^2$)\\
		\hline
		\hline
		0 (S1) & 0.6 &  0.6 & 0.024 & 0.25 \\
		0.18 & 1.9 &  1.2 & 0.021 & 0.25 \\
		0.31 &  7  & 5.5 & 0.021 & 0.25 \\
		\hline
		\end{tabular}}
		\end{center}
		\caption{The parameters of the two-band analysis of the normal state resistivity below $T$= 15 K. }
		\label{tab:2}
\end{table}
\end{center}     

\begin{figure}
\includegraphics[width=\columnwidth,clip]{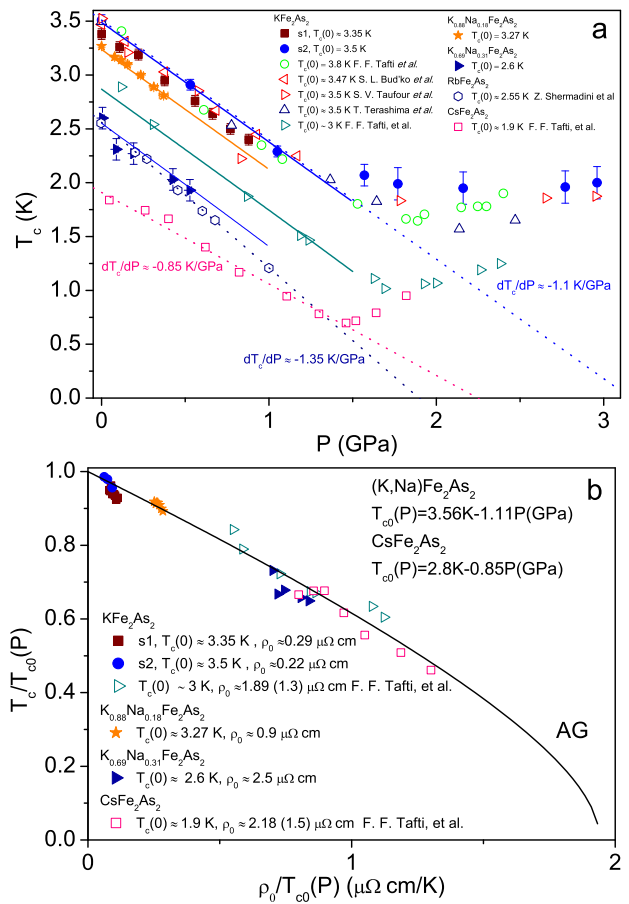}
\caption{(color online) a) The pressure dependence of
$T_c$ for K$_{1-x}$Na$_x$Fe$_{2}$As$_{2}$ 
samples from this work and those taken from 
the literature for stoichiometric K-122  
\cite{Tafti2013,Tafti2014,Taufour2014,Terashima2014,Budko2012}, 
Cs122 \cite{Tafti2014}, Rb122 \cite{Shermadini2012}. 
Dashed lines: linear fit and solid lines: results 
obtained from Eq.\ \ref{EqAG}. b) The normalized 
critical temperature $T_c/T_{c0}(P)$ vs.\ the ratio between 
the residual resistivity at zero pressure and the critical temperature for 
clean K-122 $\rho_0/T_{c0}(P)$. 
The solid line denotes the Abrikosov-Gor'kov formula. 
The $\rho_0$ values for disordered K-122 and Cs-122 samples (in 
brackets) 
are taken from Ref.\ \onlinecite{Tafti2014} 
and have been 
normalized according to procedure described in the text.}
\label{Tc_P}
\end{figure}

Fig.\,\ref{mag_all} shows the $T$ dependence of the magnetization for 
K$_{1-x}$Na$_x$Fe$_{2}$As$_{2}$ single crystals. The $T_c$ 
for the samples is defined 
using linear extrapolations as shown in Fig. \ref{mag_all}. Using this criterion 
for the critical temperature similar $T_c$ values are obtained from our 
magnetization data and from specific heat measurements performed on the 
crystals from the same batch \cite{Abdel-Hafiez2012,Grinenko2013}. This 
$T_c$ criterion results in slightly lower $T_c$ values than that 
obtained from the zero-resistivity value usually used as a criterion for $T_c$ 
in the transport measurements under pressure.\cite{Taufour2014, Tafti2013} 
However, the observed small discrepancy between these two criteria does not 
affect the main conclusion of our work. It is seen in Fig.\,\ref{mag_all} that $T_c$ of all samples monotonically decreases with 
pressure up to about 2~GPa. The sample S2 ($x$ = 0) 
was measured up to the maximum pressure of about 3~GPa Fig.\,\ref{mag_all}b. 
It is seen in Fig.\,\ref{Tc_P}a that the 
pressure dependence of $T_c$ for this sample has a shallow minimum at about 
2~GPa in accord with the data obtained in the experiments with He as a 
pressure medium.\cite{Taufour2014}

The pressure dependence of $T_c$ for our K$_{1-x}$Na$_x$Fe$_{2}$As$_{2}$ 
system and the available data in the literature for 
K-122,\cite{Tafti2013,Tafti2014,Taufour2014,Terashima2014,Budko2012} Cs-122 
\cite{Tafti2014} and Rb-122 \cite{Shermadini2012} are summarized in Fig.\,\ref{Tc_P}a. 
One can see that K$_{1-x}$Na$_{x}$Fe$_{2}$As$_{2}$ and 
stoichiometric K-122 with different $T_c$ values at zero pressure exhibit
a similar pressure derivative $dT_c/dP \approx$ -1 K/GPa. However, the 
sister compounds Rb-122 and Cs-122 show
clearly different pressure derivatives -1.35 K/GPa and -0.85 K/GPa, correspondingly.
As was shown in our previous work, the $T_c$ 
of the K$_{1-x}$Na$_x$Fe$_{2}$As$_{2}$ system at zero pressure is defined mainly 
by a single strongly coupled superconducting band. \cite{Grinenko2014} 
To demonstrate that the
Na substitution suppresses $T_c$ due to the pair-breaking effect, only, 
and that it does not affect the superconducting coupling constants 
$\lambda$ we analyzed our
experimental data adopting
the single-band Abrikosov-Gor'kov (AG) formula 
modified for the $d$-wave case \cite{Radtke1993}
\begin{equation}
-\ln \left( \frac{T_c}{T_{c0}} \right) =
\psi \left( \frac{1}{2}+\frac{\alpha T_{c0}}{2\pi T_c} \right)-\psi \left( \frac{1}{2}\right) \ ,
\label{EqAG}
\end{equation}
where $\alpha = 1/[2\tau_{\rm eff}T_{c0}$] is the pair-breaking
parameter and $1/\tau_{\rm eff} \propto \rho_0$ is the 
effective scattering rate due
to impurities created by the Na substitution.
The quantities $\tau_{\rm eff}$, and $T_{c0}$ can be pressure dependent.
Taking this into account, we have found that all available data for 
the K$_{1-x}$Na$_{x}$Fe$_{2}$As$_{2}$ system and 
the disordered K-122 crystals can be fitted by the
AG formula assuming that the critical temperature of the 
clean system (without pair-breaking impurities) varies linearly with 
pressure and can be approximated by 
$T_{c0}(P) = 3.56$~K $- 1.11P$ up to $P\approx$~1.5 GPa. 
In contrast, it is found that $\tau_{eff}$ is pressure independent. 
In general, the observed AG behavior in the 
K$_{1-x}$Na$_{x}$Fe$_{2}$As$_{2}$ system may be considered as a justification 
that Na acts as a pair-breaking impurity in agreement with Ref.\ \onlinecite{Grinenko2014}. 

The independence of $T_{c0}(P)$ on the Na concentration
is illustrated in 
Fig.\ \ref{Tc_P}b, where we plot the normalized $T_c/T_{c0}(P)$ versus the 
residual resistivity $\rho_0$ measured at zero pressure divided by $T_{c0}(P)$. 
It is seen that all available data up to 1.5~GPa scale on a
single curve. The observed scaling indicates, also, that both $\tau_{eff}$  and
$\rho_0 \propto 1/\tau_{eff}$ is nearly 
pressure independent. The insensitivity of 
$\rho_0$ to pressure for K-122 is confirmed experimentally. \cite{Tafti_p} 
Interestingly, 
the Cs-122 data can be scaled to the same curve using 
their different $T_{c0}(P) = 2.8K - 0.85P$, only. Note, that the clean limit 
$T_{c0}(P=0)=2.8$ K is rather reasonable for Cs-122 compound taking into account 
available literature data.\cite{Sasmal2008, Wang2014, Tafti2014}. For example, the reported in Ref.\ \onlinecite{Sasmal2008}  $T_c \approx$ 2.6 K is quite close to our adopted  $T_{c0}(P=0)$ value.\\   
 
\subsection{Crystal structure}

\begin{center}
\begin{table}
		\vspace{0.5cm}
		\begin{center}
		\scalebox{1}{
		\begin{tabular}{|c|c|c|c|c|}
		\hline
		 $x$ & $T$(K) & $a$ ($\r{A}$)  & $c$ ($\r{A}$) & $z_{As}$\\
		\hline
		\hline
		0.37(10) & 300 & 3.8341(2) & 13.656(2) & 0.3541(1) \\
		0.37(10) & 300 & 3.8353(3) & 13.693(2) & 0.3541(2) \\
		0.22(0.12) & 100 & 3.8172(5) & 13.596(2) & 0.35461(18) \\
		\hline
		\end{tabular}}
		\end{center}
		\caption{The crystal structure parameters of 
		K$_{1-x}$Na$_x$Fe$_2$As$_2$ single crystals. The values for 
		the partially Na substituted samples have been obtained from x-ray data. 
The $x$ values are in a reasonable agreement with EDX data (0.31 and 0.18) obtained on the crystals from the same batches.}
		\label{tab:1}
\end{table}
\end{center}

\begin{figure}
\includegraphics[width=\columnwidth,clip]{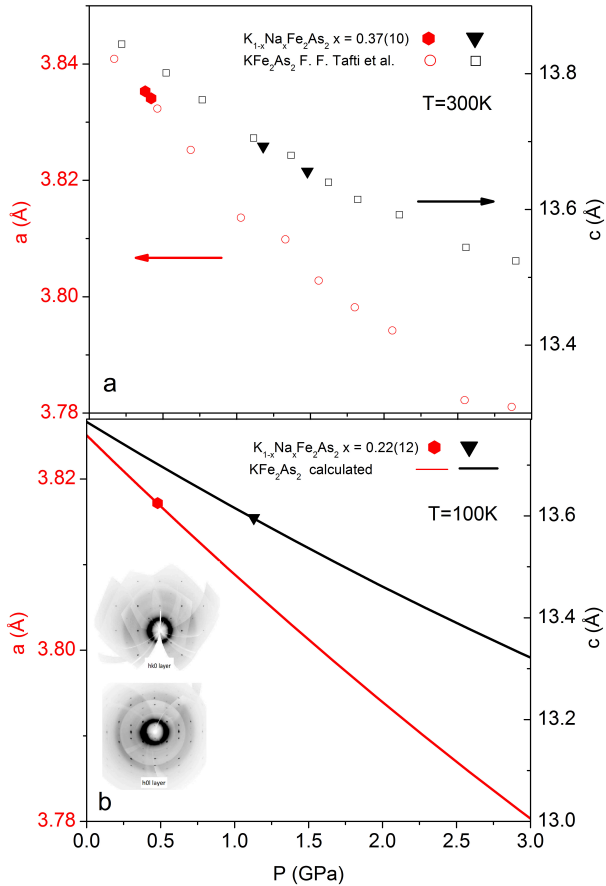}
\caption{(color online) The hydrostatic pressure dependence of 
the lattice parameters for K-122. The experimental data for K-122 are 
taken from Ref.\ \onlinecite{Tafti2014}. The data points for 
K$_{1-x}$Na$_x$Fe$_{2}$As$_{2}$ correspond to the lattice parameters which
have been obtained from x-ray data at atmospheric pressure. 
The data points are shifted along the
pressure axis to fit the values for KFe$_{2}$As$_{2}$ at 
external hydrostatic pressure.
a) The data obtained at room temperature. Close symbols - the experimental data 
for K$_{1-x}$Na$_x$Fe$_{2}$As$_{2}$, open symbols are taken 
from Ref.\ \cite{Tafti2014} b)  Close symbols - the experimental data 
for K$_{1-x}$Na$_x$Fe$_{2}$As$_{2}$ obtained at $T\approx$ 100 K. 
Solid lines - theoretical calculations corresponding to the
low-temperature regime. 
Inset: single crystalline diffraction pattern of (hk0), and (h0l) planes  
for a Na doped crystal.}
\label{c_structure}
\end{figure}

It is well known that isovalent chemical substitution in iron pnictides 
suppresses the SDW phase and can induce superconductivity 
(see for example Refs.\ \onlinecite{Walmsley2013,Rullier-Albenque2010}). 
In this case one of the relevant factors which controls 
the phase diagram is the change of 
the structure parameters \cite{Engelmann2013}. 
This effect can be considered as a chemical pressure. To understand the 
effect of the Na substitution on the lattice parameters, 
we performed 
single crystal diffraction measurements at room temperature and at $T= 100$~K. 
Selected crystallographic results are given in Table \ref{tab:1}.
The lattice parameters of K$_{1-x}$Na$_x$Fe$_{2}$As$_{2}$ 
single crystals obtained at $T$ = 300 K are compared with 
the experimental data for K-122 \cite{Tafti2014} obtained at room temperature  (Fig.\,\ref{c_structure}a)
and the lattice parameters obtained at $T$ = 100 K are compared with the results of our DFT calculations relevant for low temperature (Fig.\,\ref{c_structure}b). 
We note that the
lattice parameters only weakly depend on temperature below $T \sim$ 100 K 
\cite{Avci2012}. 
It is seen in Fig.\,\ref{c_structure} that the chemical 
pressure due to the Na substitution produces a stronger effect along 
the $c$ axis by about a factor of 2 - 3 
than in the $ab$ plane as compared to the hydrostatic external pressure effect 
on K-122.   
Taking into account that $T_c$ of K-122 is equally sensitive to uniaxial pressure applied in the $ab$ plane and along the $c$ axis ($\partial T_c/\partial P_{ab} \approx - \partial T_c/\partial P_c$) with the resulting superposed pressure derivative 
$dT_c/dP=2\partial T_c/\partial P_{ab}+\partial T_c/\partial P_c$ \cite{Budko2012}, 
we can conclude that the chemical pressure effect on 
$T_c$ due to the Na substitution is rather weak. \cite{remark_pressure} 
This provides additional support for 
our assumption that the main reason of the observed $T_c$ suppression 
in Na doped crystals is given by the 
pair-breaking effect. \\

\section{Theoretical analysis}

To get a better understanding of the pressure effect on $T_c$ 
we performed a density functional theory (DFT) based
theoretical analysis of the band structure under pressure. \\

\subsection{Computational details}

The calculations were performed using the Vienna Ab-Initio Simulation 
Package (VASP)\cite{vasp1,vasp2} within the generalized gradient approximation (GGA)\cite{gga}.
The Perdew, Burke and Ernzerhof (PBE)\cite{00004}  functional was used to calculate 
the exchange-correlation potential.
We used projected augmented-wave (PAW) pseudopotentials\cite{paw} for all the atomic 
species involved, and in order to achieve a satisfactory
degree of convergence the integrations over the Brillouin Zone (BZ) was 
performed considering 12$\times$12$\times$8 and 24$\times$24$\times$24 
uniform Monkhorst and Pack\cite{kpoints} grids for the conventional (4 K, 8 Fe and 8 As atoms)
and single formula unit (1 K, 2 Fe and 2 As atoms) cells, respectively,
and energy cutoff up to 500 eV. The Fermi surface cuts of chosen lattice 
structures (see below) as a function of $k_z$ have been computed
using a $64\times64\times4$ k-points grid.\\

\subsection{Band structure and Fermi surfaces}

\begin{center}
\begin{table}
		\vspace{0.5cm}
		\begin{center}
		\scalebox{1}{
		\begin{tabular}{|c|c|c|c|c|c|}
		\hline
		 {}& $a $ (\AA) & $b $ (\AA) & $c $ (\AA) & z$_{As}$ \\
		\hline
		\hline
		EXP & 3.83 & 3.83 & 13.79 & 0.353 \\
		PM & 3.79 & 3.79 & 13.99 & 0.346 \\
		\hline
		\end{tabular}}
		\end{center}
		\caption{Experimental \cite{Avci2012} and paramagnetic fully optimized structural parameters 
		of KFe$_2$As$_2$. 
}
		\label{tab:3}
\end{table}
\end{center}

\begin{figure}
\begin{center}
\includegraphics[width=\columnwidth,clip]{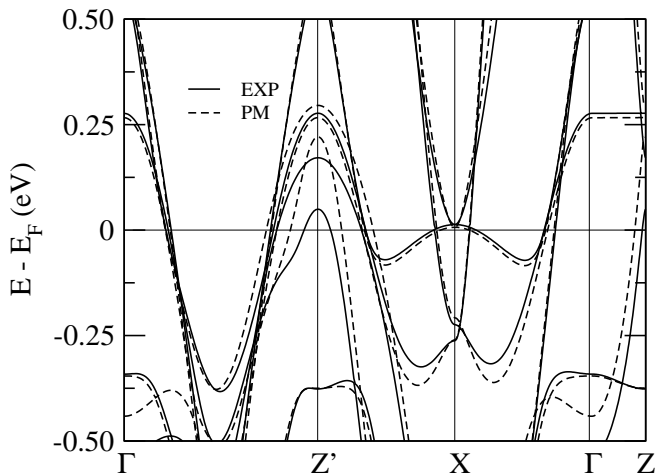}
\caption{KFe$_2$As$_2$ band structure considering different crystal lattice
 parameters (namely, the experimental and PM ones, see table \ref{tab:3}).}
\label{bandstr}
\end{center}
\end{figure}

\begin{figure}
\begin{center}
\includegraphics[width=\columnwidth,clip]{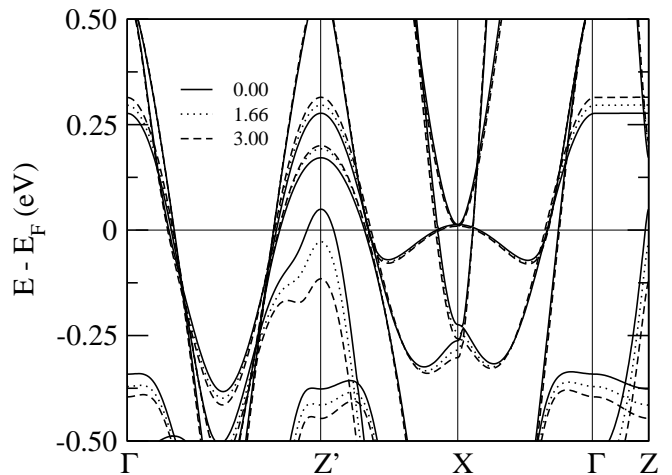}
\caption{KFe$_2$As$_2$ band structure for the experimental lattice structure 
(taken as $P = 0$ GPa geometry) and lattice parameters extrapolated with compressibility at 1.66 and 3.00 GPa (see Tab. \ref{tab:1a}).}
\label{Bands_pressure_PM}
\end{center}
\end{figure}

\begin{figure}
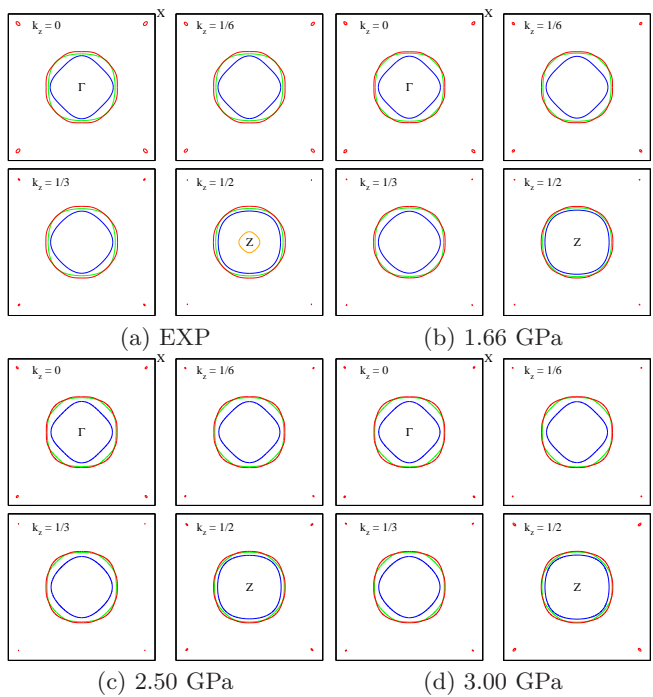

\begin{tabular}{cc}
\includegraphics[width=10pc,clip]{FS_cuts_EXP_450cutoff.eps} & \includegraphics[width=10pc,clip]{FS_cuts_166GPa_450cutoff_compress.eps} \\
(a) EXP &  (b) 1.66 GPa\\
\includegraphics[width=10pc,clip]{FS_cuts_250GPa_450cutoff_compress.eps}& \includegraphics[width=10pc,clip]{FS_cuts_3GPa_450cutoff_compress.eps}\\
(c) 2.50 GPa & (d) 3.00 GPa\\
\end{tabular}
\caption{(color online) Fermi surfaces cuts along the positive $k_z$ direction for KFe$_2$As$_2$ for the experimental crystal structure and for the extrapolated lattice parameters using the predicted compressibility. The inner, middle and outer FS's around  the $\Gamma$-point are identified as $\alpha$, $\zeta$ and $\beta$, respectively. (For the orbital content of FS's see text.)}
\label{FS_exp}
\end{figure}

We studied the influence of the structural properties on the electronic paramagnetic (PM) band structure of KFe$_2$As$_2$.
In Fig. \ref{bandstr}, we report the calculated band structure considering the experimental \cite{Avci2012} and the optimized lattice constants in the PM phase. The predicted (theoretical) and experimental lattice constants are compared in table \ref{tab:3}. In spite of relatively small differences between experimental and theoretical lattice parameters (in a range of 1 - 2 \%), 
the Fermi surface cross-sections calculated with experimental lattice parameters reveal important differences with 
respect to the theoretical calculated ones.  
The radii of the $\Gamma$-centered hole Fermi surfaces, namely $\alpha$, $\zeta$ and $\beta$, do not change sensibly, while the small electron-pockets around the X point shrink considering the theoretical lattice constants. The small (but irrelevant) electron pocket at Z is found to become larger using the theoretical lattice parameters.
Since the discrepancy between theoretical and experimental crystal lattices affects bands,
in the further analysis we stick to the experimental structure  \cite{Avci2012}: we consider this one as the reference geometry at $P=0$, {\em i. e.}, $z_{As}^{exp}= z_{As}(P=0)$, $a^{exp}=a(P=0)$, and $c^{exp}=c(P=0)$. Then, we extracted the lattice parameters calculating the compressibility for the $z_{As}^{th}(P)$, $a^{th}(P)$ and $c^{th}(P)$ by first principles as a function of the pressure.
The lattice constants, 
obtained in this way, are given in Tab. \ref{tab:1a}. These data was also used for the analysis of the chemical pressure (see Fig. \ref{c_structure}b). The calculated band structure and Fermi surface cross-sections with these lattice parameters are given in Figs. \ref{Bands_pressure_PM}, and \ref{FS_exp} for different pressures. We underline that experiments found the two outermost FSS around the $\Gamma$ point not intersecting\cite{Terashima2013}, however, DFT-GGA find them crossing. Keeping into account this effect on the band structure, the orbital character of the Fermi Surface is the following:
{\em i)} the main contribution to the $\alpha$ FS is from $d_{z^2}$ (along the $\Gamma$Z direction) and from $d_{xz+yz}$ (along the $\Gamma$X direction) orbitals;
{\em ii)} we call $\zeta$ the band with character $d_{xy}$ (the green FS along the $\Gamma$Z direction which becomes red along the $\Gamma$X, see Fig. \ref{FS_exp});
{\em iii)} in the same way, we refer to $\beta$ the FS with $d_{xz+yz}$ character (the outermost red FS along the $\Gamma$Z direction becoming the middle green one along $\Gamma$X, see Fig. \ref{FS_exp});
{\em iv)} the main contribution to 
the lobes $\epsilon$ near the X point is from $d_{xz+yz}$ and $d_{xy}$;
{\em v)} the character of the three dimensional FS at the Z point (orange band in Fig. \ref{FS_exp}) is $d_{xy}$.

In order to understand the origin of the reversed behavior of $T_c$ under
pressure, first, we compared the band structure at ambient ($P$ = 0 GPa), 
close to the critical pressure ($P$ = 1.66 GPa) and at high pressure ($P$ = 3 GPa), shown in Fig.\ \ref{Bands_pressure_PM}. The shape of the relevant Fermi 
surfaces are not strongly affected in this range of pressure, indicating 
that the origin of the non-monotonic pressure dependence of $T_c$ is not 
due to changes of the Fermi surface topology in accord with experimental 
findings. \cite{Terashima2014, Tafti2013} 

\begin{center}
\begin{table}
		\vspace{0.5cm}
		\begin{center}
		\scalebox{1}{
		\begin{tabular}{|c|c|c|c|c|}
		\hline
		 P (GPa) & $a $ (\AA) & $c $ (\AA) & $z_{As}$ & $z_{As}$ (\AA)\\
		\hline
		\hline
		0.00 & 3.83 & 13.79 & 0.353 & 1.42 \\
		1.00 & 3.81 & 13.62 & 0.355 & 1.42 \\
		1.66 & 3.80 & 13.51 & 0.356 & 1.43 \\
		2.00 & 3.80 & 13.45 & 0.356 & 1.43 \\
		2.50 & 3.79 & 13.37 & 0.357 & 1.43 \\
		3.00 & 3.78 & 13.28 & 0.358 & 1.43 \\
		4.00 & 3.76 & 13.11 & 0.359 & 1.43 \\
		\hline
		\end{tabular}}
		\end{center}
		\caption{Structural parameters for the four different pressure values (P = 0 GPa is relative to the
		experimental cell taken as reference, see text). }
		\label{tab:1a}
\end{table}
\end{center}

\section{Discussion}
\begin{figure}
\begin{center}
\includegraphics[width=\columnwidth,clip]{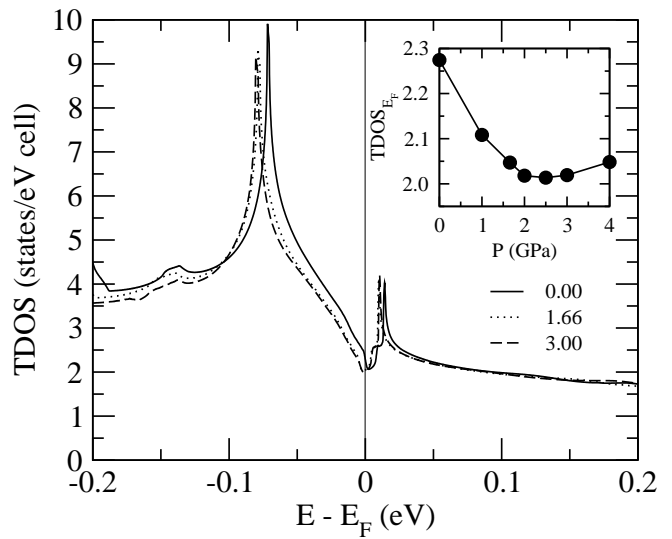}
\caption{The variation of the total density of states (TDOS) under 
pressure for KFe$_2$As$_2$ (per one spin). Inset: pressure dependence of the TDOS at the Fermi level.}
\label{dos}
\end{center}
\end{figure}

The total density of states (TDOS) for different values of pressure is presented in Fig.\ \ref{dos}.  At zero pressure TDOS has a minimum at energies of a few meV above the Fermi level, which is followed by a peak in TDOS. Applying  pressure leads to a redistribution of the states, that formally looks like Fermi level moves up through the minimum towards the sharp peak. As a result TDOS at the Fermi-level first decreases with the pressure, has a minimum at the pressure of about 2~GPa and increases back as shown in the inset of Fig.\ \ref{dos}.

The change of the DOS may affect $T_c$. As it was shown from thermodynamics, in the clean limit the $T_{c0}$ of K-122 is determined
mainly by a single band.\cite{Grinenko2014,Reid2012, Hardy2013a} 
For further discussion we adopt the model proposed in Ref.\ \onlinecite{Grinenko2014} with coupling constants $\lambda_z \approx 0.9$ and $\lambda_{\phi} \approx 0.8$.  
In this case the critical temperature can be written as 
$ln(\omega_b/T_{c0})=(1+ \lambda)/a \lambda$, where $\omega_b$ is the characteristic bosonic frequency of the superconducting "glue", $\lambda = \lambda_z \approx 0.9$ is the coupling constant and $a=\lambda_{\phi}/\lambda_z \approx 0.9$.
Then, assuming that $\omega_b$ and $a$ are pressure independent, for the pressure dependence of $T_c$ we have
\begin{equation}
T_{c0}(P) \approx T_{c0}(0)\frac{exp(-\frac{1+\lambda(P)}{a\lambda(P)})}{exp(-\frac{1+\lambda(0)}{a\lambda(0)})}.
\label{Tc_P_Eq}
\end{equation}
Taking into account that $\lambda=gN_i$, with $g$ as 
the effective interaction and $N_i$ as the
partial density of states (PDOS) of the leading superconducting band, then for 
small variations of $g$ and $N$ the variation of the critical 
temperature under pressure can be written as
\begin{equation}
 \frac {\delta T_{c0}}{T_{c0}} \approx \frac {1}{a\lambda}\left[\frac {\delta g}{g}+\frac {\delta N_i}{N_i}\right].
\label{Tc_var}
\end{equation} 

\begin{figure}
\begin{center}
\includegraphics[width=\columnwidth,clip]{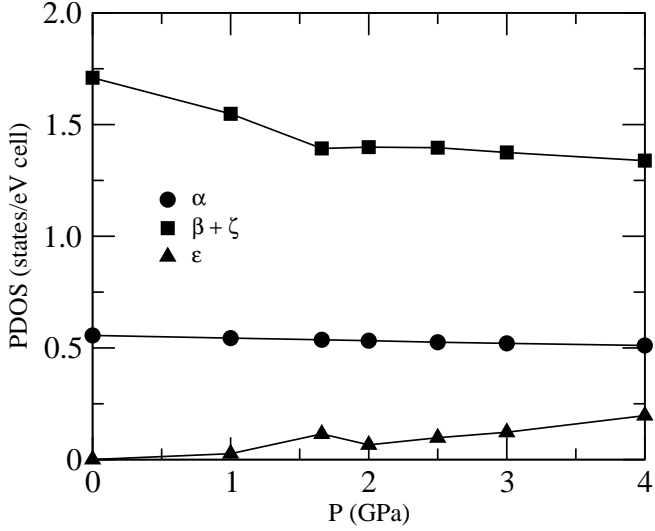}
\caption{The variation of the band-resolved DOS at the Fermi level under 
pressure for KFe$_2$As$_2$.}
\label{pdos}
\end{center}
\end{figure}

\begin{figure}
\begin{center}
\includegraphics[width=\columnwidth,clip]{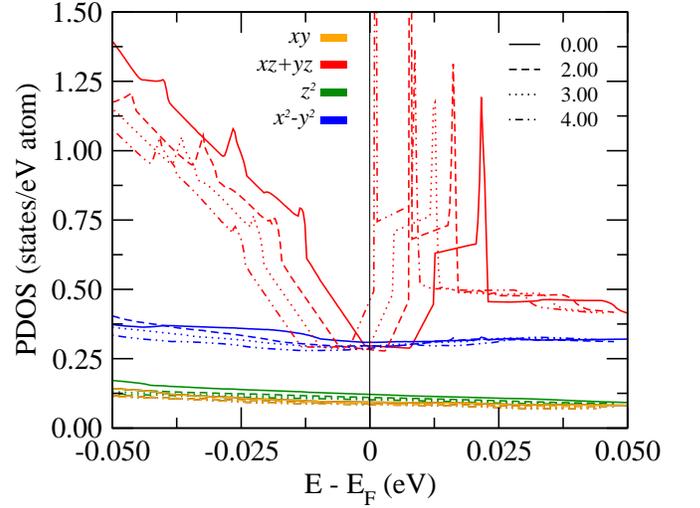}
\caption{The variation of the various Fe 3d orbitals projected
density of states under pressure for KFe$_2$As$_2$ per 
spin and one Fe atom.}
\label{pdos_orbit}
\end{center}
\end{figure}

As shown in Fig.\ \ref{pdos} the partial density of states (PDOS) of the 
various FSSs 
behave differently with pressure. $N_{\alpha}(0)$ weakly 
depends on 
pressure up 4~GPa while 
$N_{\beta+\zeta}(0)$ 
decreases from 0 to 1.66~GPa by
about 20\% and then remains practically constant. At this point, it is 
clear that the weak increase of the TDOS
above 2~GPa (Fig.\ \ref{dos}) comes mainly from the $\epsilon$ band. 
Indeed, the $N_{\epsilon}(0)$ increases up to 4~GPa as it can be seen in Fig.\ \ref{pdos}. 
The non-monotonic pressure dependence of 
$N_{\beta+\zeta}(0)$ resembles the experimentally observed $T_{c0}(P)$ dependence (Fig.\ \ref{Tc_P}). 
Using Eq.\ \ref{Tc_P_Eq} one can estimate from 
$N_{\beta+\zeta}(0)$ that it corresponds to 
a decrease of the critical temperature from $T_c(P=0) \approx 3.5$~K down 
to $T_c(P=1.66$ GPa) $\sim$ 2.5~K assuming that $g$ is pressure independent. 
However, within a more quantitative study the change of the pairing interaction 
$g$ (see Eq.\ \ref{Tc_var}) and of the spin-fluctuation spectrum, i.e, $\omega_b$, 
under pressure should be taken into account, too.  Above $P=1.66$ GPa 
$T_c$ would remain
nearly constant. This is consistent with a broad minimum in the pressure 
dependence of $T_c$ reported in Refs.\ \onlinecite{Taufour2014, Terashima2014} 
and observed in our data, too (see Fig.\ \ref{Tc_P}). 
Note, that due the overlap of the $\beta$ (mainly of $xz+yz$ orbital 
content) and $\zeta$ (mainly of $xy$ orbital content) FSSs we could not 
separate them unambiguously. However, as shown in Fig.\ \ref{pdos_orbit} the orbital-projected PDOS for all bands except $xz+yz$ decreases in the whole pressure range.
Therefore, 
based on our theoretical analysis we 
suggest that the experimentally observed 
non-monotonic pressure dependence of $T_c$ is 
related to the $xz+yz$ band. Then, the $\beta$ FSS, predominantly having 
$xz+yz$ character (experimentally the
middle $\zeta$ FSS \cite{Terashima2013}), is responsible for the
superconductivity in the K-122 compound. 
We note, that the presence of a
nodal SC gap of about 0.7 meV (as
estimated from our specific heat data \cite{Grinenko2014}) on the 
middle FSS is in a quantitative agreement with the 
laser ARPES experiments. \cite{Okazaki2012} However, the observed very large 
nodeless SC gap on the inner FSS \cite{Okazaki2012} disagrees with the specific heat data. 
Therefore, further investigations are required to understand this discrepancy.

\begin{table*}
\begin{center}
\begin{tabular}{|c|c|c|c|c|c|c|c|c|c|c|c|c|c|c|} \hline
{} & {} & \multicolumn{5}{c|}{FS area (BZ \%)} & \multicolumn{5}{c|}{FS frequencies (kT)} & \multicolumn{3}{c|}{$m^*/m_e$} \\
\hline 
Band & $k_z$ & dHvA$^{a,c}$ & A.$^{a}$& A.$^{b}$& LDA$^a$ & GGA & dHvA$^{c}$ & QOM$^{d}$ & LDA$^{e}$ & DMFT$^{f}$ & GGA & dHvA$^{c}$ & A.$^{a}$ & GGA \\ 
\hline
\multirow{2}{*}{$\alpha$} & $\Gamma$ & 8.2 & 9.1 & \multirow{2}{*}{7} & 12.2  & 12.7 & 2.30 & 2.31 & 3.42 & 4.05 & 3.56 & 6.0 & 5.1 & 1.43\\
 & Z & 8.6 & 9.8 & & 13.8 & 15.0 & 2.39 & 2.39 & 3.86 & 4.20 & 4.20 & 6.5 & 6.6 & 2.92\\
\hline
\multirow{2}{*}{$\zeta$} & $\Gamma$ & 10.3 & 12.2 & \multirow{2}{*}{-} & 16.7 & 17.5 & 2.89 & 2.89 & 4.67 & 0.94 & 4.93 & 8.5 & 11.0 & 2.43\\
 & Z & 15.7 & 17.0 & & 17.4 & 17.8 & 4.40 & 4.40 & 4.88 & 3.25 & 5.01 & 18.0 & 17.7 & 2.47\\
\hline
\multirow{2}{*}{$\beta$} & $\Gamma$ & 25.6 & 27.3 &\multirow{2}{*}{22} & 20.8 & 19.2 & \multirow{2}{*}{7.16} & \multirow{2}{*}{7.18} & 5.82 & 6.62 & 5.39 &  \multirow{2}{*}{19}& 16.3 & 2.42 \\
 & Z & - & 30.0 & & 21.6 & 19.2 & & & 6.03 & 6.81 & 5.40 & & 17.9 & 2.35\\
\hline
\multirow{2}{*}{$\epsilon$} & $\Gamma$ & 0.86 &  \multirow{2}{*}{2.1} & \multirow{2}{*}{2.5} & 0.36 & 0.157 & 0.24 & 0.24 & 0.03 & - & 0.02 & 6.0 & 5.6 & 0.30 \\
 & Z & 1.29 & & & 0.11 & 0.003 & 0.36 & 0.36 & 0.10 & 0.42 & 0.03 & 7.2 & 4.1 & 0.41\\
\hline
\end{tabular}
\caption{Cross sectional areas, dHvA frequencies and effective masses obtained by different experimental and theoretical techniques:
A.= ARPES, QOM= Quantum Oscillations in Magnetostriction. The areas are expressed in term of the \% area occupied of the 2D BZ. The dHvA frequencies are expressed in kT. Our calculations are shown in the ``GGA'' columns. 
$^a$Ref. \onlinecite{Yoshida2012}, $^b$Ref. \onlinecite{Sato2009},
$^{c}$Ref. \onlinecite{Terashima2013}, $^{d}$Ref. \onlinecite{Zocco2014}, $^e$Ref. \onlinecite{Terashima2010}, $^{f}$Ref. \onlinecite{Backes2014}.
Note, that strong electronic correlations are neglected in LDA and GGA calculations. These correlations presumably cause the inversion of $\beta$ and $\zeta$ bands in the experiment, see for example Refs.\ \onlinecite{Kotliar2011, Backes2014}.} 
\label{tab:4}
\end{center} 
\end{table*}

\begin{table}
\begin{center}
\begin{tabular}{|c|c|c|c|c|c|} 
\hline 
Band & Pos. & 0 GPa & 1.66 GPa & 2.50 GPa & 3.00 GPa \\ 
\hline
\multirow{2}{*}{$\alpha$} & \multirow{2}{*}{inner} & 3.56 & 3.54 & 3.49 & 3.47 \\
 &  & 4.20 & 4.58 & 4.51 & 4.60\\
\hline
\multirow{2}{*}{$\zeta$} & \multirow{2}{*}{middle} & 4.93 & 5.16 & 5.27 & 5.31 \\
 &  & 5.01 & 5.24 & 5.35 & 5.38\\
\hline
\multirow{2}{*}{$\beta$} & \multirow{2}{*}{outer} & 5.39 & 5.49 & 5.56 & 5.59\\
 &  & 5.40 & 5.51 & 5.60 & 5.66\\
\hline
\end{tabular}
\caption{Low (top row) and high (bottom row) dHvA frequencies (kT) as a 
function of the pressure for different FS bands. The band position is indicated in the second column.} 
\label{tab:5}
\end{center} 
\end{table}

In this context, the observed AG-type scaling of the 
pressure dependencies of $T_c$ for nodal K,Na-122 and Cs-122 suggests that these compounds have the same superconducting gap symmetry. However, the Cs-122 and the (K,Na)-122 systems 
have different $T_{c0}$ and hence different coupling constants.
In the case of extended 
$s$-wave pairing symmetry such as $s\pm$ -wave SC with accidental nodes,  
the nodal position 
is not fixed on the Fermi surface \cite{Reid2012,Mishra2009,Wang2013}. Therefore, 
nodes can be lifted by impurities, pressure or isovalent substitution. 
In some cases a
reentrant behavior of the nodes can be observed.\cite{Wang2013} 
In contrast, for K,Na-122 single crystals we observed
a monotonic increase of the residual electronic specific heat $C_{el}/T$ 
with Na substitution and a
nearly Na independent $T^2$ behavior of $C_{el}$ at 
low-temperature. \cite{Abdel-Hafiez2013,Grinenko2014} Therefore, our 
data are consistent with a
symmetry protected $d$-wave SC in these compounds 
which is stable at least up to 
the pressure where $T_c$ has a minimum (see Fig.\ \ref{Tc_P}b). 

Finally, we compare the calculated Fermi surface cross sections 
with the experimental ones obtained from dHvA measurements \cite{Terashima2014}. 
These dHvA experiments show, that the size
of the lobes near the X-point decreases with pressure and 
that of the $\alpha$ cylinder weakly increases up to 2.47~GPa, 
the maximum pressure reached experimentally. The trend in the calculated Fermi surface 
cross sections for this pressure range, i.e,\ up to 2.5~GPa (see Fig.\ \ref{FS_exp}) 
is consistent with the 
corresponding dHvA frequencies. However, increasing the pressure above 
2.5~GPa, e.g.\ in the case of 3~GPa shown in Fig.\ \ref{FS_exp}(d), 
the size of the lobes maintain their dimension. Thus, this prediction 
could be verified experimentally by studies around this pressure. 

For this purpose, in Tab.\ \ref{tab:4} we collected the calculated 
cross section of various
FS bands, dHvA frequencies and the $m^{*}/m_e$ ratios as found in 
the literature from both measurements and simulations.
Together with them, we add our GGA results calculated for
the experimental crystal structure. The dHvA frequencies and the $m^{*}/m_e$-ratio 
have been estimated calculating the extremal FS areas and the relative orbit 
frequencies using its implementation in the SKEAF code\cite{Skeaf}.
The GGA calculated FS cross sections are in qualitative agreement with 
respect to the experiments (except $\epsilon$ FS). However, the theoretical $m^{*}/m_e$ ratios essentially  underestimate experimental values pointing to sizable correlation effects in accord with a previous study.  
In this context we note that the observed large difference of the effective 
resistivity coefficients $A_1$ and $A_2$ discussed above (see Tab.\ref{tab:2}),
is likely to be caused by the high-energy mass renormalizations and
not related directly to the pairing interactions. This suggestion
is also consistent with the significant pressure dependence of the 
approximate effective single $A$ coefficient reported by V.\ Taufour {\it et al.} \cite{Taufour2014}.

At this point, we extended the study to high pressure and we report the 
results of calculated dHvA frequencies and we compared them with 
experimental measurements.\cite{Terashima2014} The results obtained in 
the calculations are shown in Tab.\ \ref{tab:5} for the FS's shown in 
Fig.\ \ref{FS_exp}. We observed that the largest dHvA frequency increases 
for the $\alpha$-band as a function of the pressure, in agreement with 
experiment.\cite{Terashima2014} At the same time, the lowest frequency slightly decreases with pressure  
overestimating the increase of the three dimensionality of the 
electronic structure under pressure as compared to experiment.\cite{Terashima2014} 

We confirm, moreover, the absence of a Lifshitz transition under pressure. \cite{Lifshitz_expl} Note, that the  variation of the TDOS at the Fermi level in the pressure range up to 3 GPa is below 15\% as shown in Fig.\ \ref{dos}. 
Such a relatively small variation of the TDOS for multiband systems 
can be hardly distinguished in experiments such 
as transport measurements. Therefore, we suppose that the observation of  
a nearly pressure independent Hall coefficient and residual resistivity  \cite{Tafti2013,Tafti_p} cannot exclude such relatively small variations of the DOS.\\

\section{Conclusion}

In conclusion, we investigated the effect of 
hydrostatic pressure and Na substitution on the superconducting and the 
normal-state properties of KFe$_2$As$_2$. We have found that the Na substitution 
noticeably affects the lattice parameters which can be considered as a 
non-hydrostatic chemical pressure. However, the comparison of the external 
hydrostatic pressure with the chemical pressure shows
that the total effect of the structural 
changes on $T_c$ due to Na substitution is rather weak. Therefore, the 
main mechanism of the $T_c$ suppression in Na doped crystals is a weak 
pair-breaking effect caused by disorder. The Na substitution enhances the residual 
resistivity and leads to a formal non-Fermi liquid behavior in the 
temperature dependence of the resistivity. We have shown that this deviation 
from the $T^2$ behavior can be described
by multiband effects due to different effective quasi-particle 
masses for different bands. The pressure dependencies of $T_c$ for 
the K$_{1-x}$Na$_x$Fe$_2$As$_2$ system can be described 
by the single-band Abrikosov-Gor'kov-type scaling curve using the Na independent 
critical temperature in a clean limit $T_{c0}$(P). These observations additionally confirm that 
the $T_c$ suppression for Na substitution is mainly caused by the pair-breaking effect.  
 The pressure dependence of $T_c$ for  CsFe$_2$As$_2$  can be also scaled on the same curve, 
 however, with a different  $T_{c0}$(P). These results suggest the same pairing symmetry 
 for this compound as for the K$_{1-x}$Na$_x$Fe$_2$As$_2$ system. Additionally, 
 we performed theoretical investigations of the band structure of KFe$_2$As$_2$ under pressure within the 
 generalized gradient approximation. 
Our results suggest, that the observed pressure dependence of $T_c$ can 
be explained by the non-monotonic variation of the PDOS 
of the $xz+yz$ derived band under 
pressure without a change of the pairing symmetry. This  $xz+yz$ 
derived band is supposed to be most relevant for 
the superconductivity of the K$_{1-x}$Na$_x$Fe$_2$As$_2$ system.\\        

\section*{Acknowledgment}

This work was supported by the EU-Japan project
(No.\ 283204 SUPER-IRON), and the DFG through the SPP 1458, the
E.-Noether program (WU 595/3-1 (S.W.)). S.W.\ thanks the
BMBF for support in the frame of the ERA.Net RUS project FeSuCo
No.\ 245, E.A.\ thanks the DFG-GRK1621 and I.M. thanks funding from RFBR (12-03-01143-a). We acknowledge D.\ Evtushinsky, K.\ Iida, K.\ Nenkov, C.\ Hess, and A.\ Yaresko for fruitful discussion.


\begin{thebibliography}{100}
\bibitem{fn_pressure} Details about the pressure cell construction will be published elsewhere.
\end{thebibliography}
\end{document}